\documentclass[11pt,a4paper]{article}
\usepackage[T1]{fontenc}
\usepackage[utf8]{inputenc}
\usepackage{lmodern}
\usepackage{amsmath,amssymb}
\usepackage{graphicx}
\usepackage{booktabs}
\usepackage{multirow}
\usepackage{siunitx}
\usepackage{microtype}
\usepackage[hidelinks]{hyperref}
\usepackage[margin=2.5cm]{geometry}
\usepackage{authblk}
\usepackage{caption}
\usepackage{subcaption}
\usepackage{siunitx}
\usepackage[backend=biber, style=numeric]{biblatex} 

\title{Cosmic muon arrival directions as a source of entropy}

\author[1]{Deepak Samuel}
\affil[1]{Central University of Karnataka, India}
\date{July 2026}

\begin{document}

\maketitle

\begin{abstract}
We demonstrate, for the first time, that the arrival directions of cosmic-ray muons from a muon telescope provide a measurable source of physical entropy. Using six months of data collected from a muon telescope, the local zenith and azimuth angles were used for randomness extraction. The result of this study show that each cosmic muon track from our detector can generate upto 4 bits with an entropy of 0.999182 bits/bit with a maximum bit rate of roughly 650 bits/s which can possibly be enhanced by improving the detector size and angular resolution. The suggested technique can be potentially used as a random number generator required for closing the settings-independence loophole in the tests for violations of Bell's inequality. The data acquisition, data reduction, and conditioning algorithms are discussed in detail.


\end{abstract}

\section{Introduction}
\label{sec:intro}

Physical entropy sources, for instance,  radioactive decay,  quantum optical
processes and thermal noise, convert unpredictable natural phenomena into digital
randomness. Cosmic-ray muons arrive at the Earth's surface with rates, energies, and directions determined by hadronic air showers and geomagnetic propagation. Although the distribution of cosmic muon physical parameters is well characterised, the intrinsically stochastic nature of their detection times and trajectories of individual muons make them natural candidates for randomness generation.

Previous cosmic-muon-based randomness studies exploited the arrival time of cosmic muons as entropy sources \cite{9159728}. Tracking detectors, such as the one reported here, provide the arrival direction in terms of the zenith and azimuthal angles, $\theta$ and $\phi$ as additional degrees of freedom. In this study, we demonstrate that these parameters can also serve as a useful source of entropy using a detector that was originally envisioned to be used as a test bench for the development of a large-scale neutrino detector.

\section{Background}
\subsection{Cosmic muon production}
Primary cosmic rays are high-energy particles, including mostly protons and alphas, that originate from sources outside the Earth, such as supernovae. Upon impact with the Earth’s atmosphere, these particles create secondary particles, such as pions, which, through subsequent decays, lead to muons, electrons, and neutrinos. Muons and neutrinos are the most abundant secondary particles on the Earth’s surface. The mean energy of cosmic muons on the Earth’s surface is approximately 4 GeV, with an intensity of approximately 1 \text{ cm}$^{-2}$\text{ min}$^{-1}$ \cite{ParticleDataGroup:2026aaa}. Additionally, muons arriving from large zenith angles are depleted because of the longer path length to the detectors, due to which they can decay to electrons and neutrinos before reaching the detectors. Therefore, the vertical flux is typically more pronounced than the horizontal flux. Seasonal variations also affect the flux, as temperature fluctuations can lead to expansion or contraction of the atmosphere, consequently leading to changes in the density \cite{abubakar2026explanation}. In denser media, the probability of interactions increases, thereby reducing the flux. Many experiments use secondary muons and neutrinos as particle sources for physics studies. For instance, the GRAPES-3 experiment focuses on the origin of very high-energy muons (\SI{1e14}{\electronvolt}) and their acceleration mechanism \cite{GUPTA2005311}. The Super-Kamiokande experiment won the 2015 Nobel Prize for the discovery of atmospheric neutrino oscillations \cite{PhysRevLett.81.1562}.
\subsection{Previous studies}
Random numbers are crucial for several cryptographic applications. Although pseudo-random number generators (PRNG) offer a fast solution, they are inevitably deterministic in nature \cite{PhysRevE.72.016220}. The advancement in quantum technologies implies a threat to conventional cryptographic protocols, necessitating high-quality randomness through True Random Number Generators (TRNG) in the post-quantum era \cite{shor1999polynomial}\cite{chen2016report}. Quantum Random Number Generators (QRNG) are a class of TRNGs that generate random numbers from quantum phenomena. A QRNG based  on two-photon quantum interference using a beam splitter was reported in \cite{Kwon:09}. In this technique, a pair of photons are sent through a beam splitter and a \textit{1} or \textit{0} is generated based on the path these photons take when they exit the beam splitter. Quantum principles dictate that the probability of the photons to take one of the two exit paths is 50\% and therefore the highly nondeterministic nature of bit patterns can be used as a source of randomness. In another QRNG, the dark noise from a silicon photomultiplier (SiPM) was used to generate random numbers. The non-deterministic arrival times of these noise pulses was measured using a high-resolution Time-to-Digital Converter (TDC). The timestamps thus measured were then used as the source of randomness. In another variation of this SiPM-based technique, it was shown that the counting of the number of photons emitted by a laser source using a SiPM as a multiphoton counter presents an useful way of generating randomness \cite{CACCIA2020164480}.
Related to our work is the study by H. Gamil et al. in which plastic scintillators coupled to SiPMs record the time of arrival of cosmic muons. The timing signals are accumulated for certain a period of time after which they are conditioned using a hashing algorithm to generate random numbers. While our study also uses cosmic muons, we report for the first time, the usage of arrival directions instead of the arrival times as source of randomness. This is based on the general understanding that the arrival direction of one muon is completely independent of the next one. 

\subsection{A random number generator for closing the setting-independence loophole}
Tests of Bell's inequality provide a stringent means to prove the validity of quantum mechanics. However several loopholes have to be addressed to rule out alternate local-realist theories \cite{kaiser2020tackling}. For instance, if communication between detecting elements in the experiment is  possible by some means, local realism can still be accounted for in the results. This is often referred as the locality loophole. The detection loophole or the fair sampling loophole is one in which the detector effects like inefficiencies can cause measurements that appear to be consistent with quantum mechanics but are actually due to local realism. Finally, the settings-independence or freedom-of-choice loophole is one in which the detector settings have a correlation with the particle properties at the source. Closing the settings-dependence loophole requires that the detector settings are arbitrarily chosen while the entangled particles are in flight. Most experiments use QRNGs for selecting the detector settings. A proposal also suggests the use of cosmic photons, specifically produced by quasars from opposite sides of the sky, to generate random bits to choose the detector settings. Violation of Bell's inequality would then require a local realist explanation implying a correlation that was placed billions of years ago \cite{PhysRevLett.112.110405}.  A similar argument can also be placed about the directions of cosmic muons which are produced from primary cosmic rays like protons that originate from galactic and extragalactic sources. To achieve a correlation between particle properties and the direction of two muons coming from opposite directions of the sky would necessitate a correlation to be placed between the sources that are far away and in the complex sequence of the decay chain from which the muon was produced. A detailed study in this aspect is outside the scope of this study but we merely state that a random number generator based on the arrival direction of the muons would be an option if such experiments are to be performed. 


\section{Detector setup}
\begin{figure}
    \centering
        \includegraphics[clip, trim=0.0cm 0cm 0.0cm 0.0cm, width=1\linewidth, page=3]{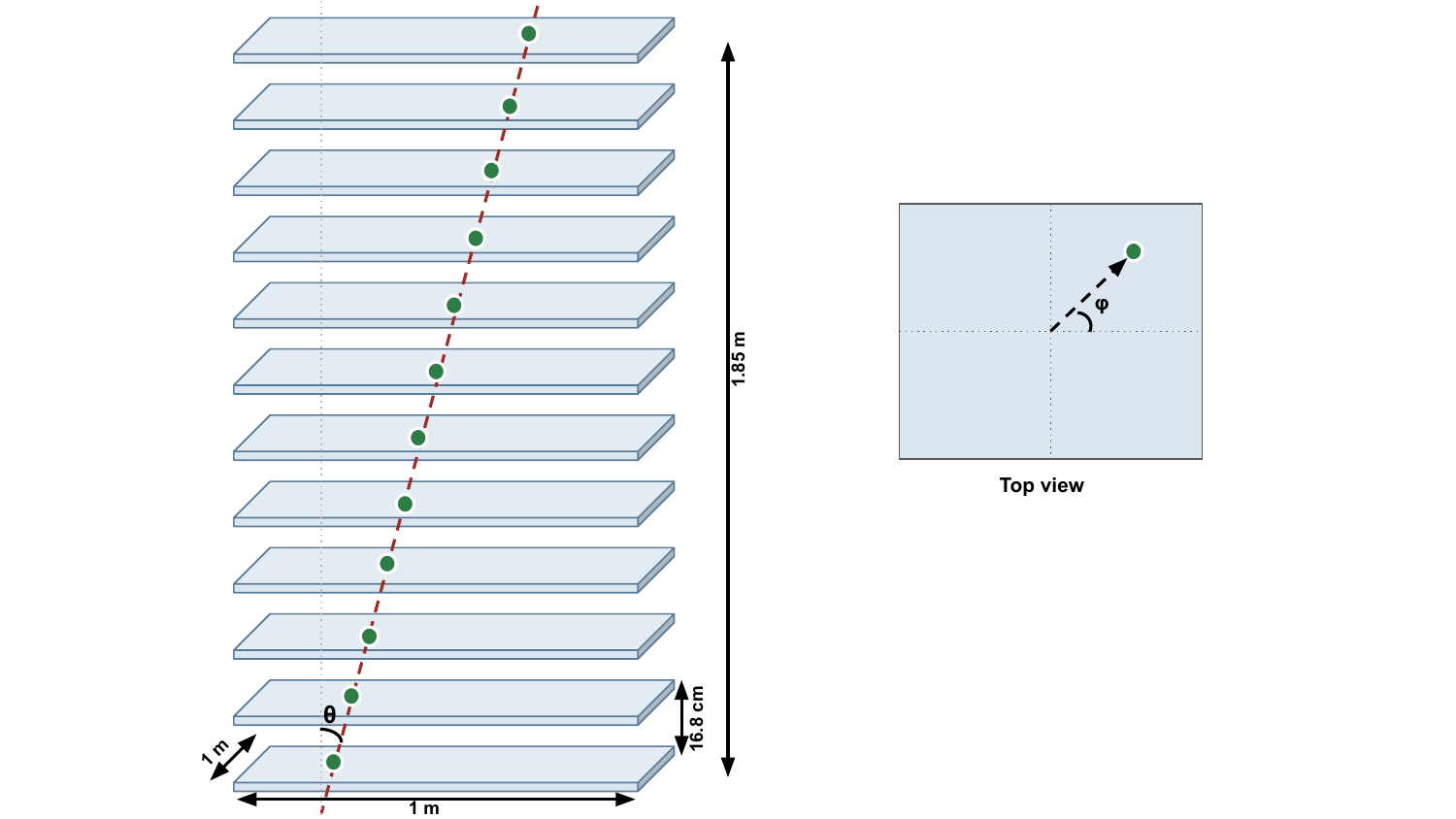}
    \caption{Schematic of the RPC stack at TIFR used for data collection reported in this study. $\theta$ is the zenith angle of the track and $\phi$, the azimuth. The dots indicate the hit positions on the RPCs.}
    \label{fig:stack}
\end{figure}

The India-based Neutrino Observatory (INO) was a proposed project aimed at measuring neutrino oscillation parameters using a 50 kton Iron Calorimeter with Resistive Plate Chambers (RPCs) as active detectors \cite{ICAL:2015stm}. Neutrinos interacting with iron layers produce charged muons that are tracked by the RPCs. The tracks are then used to reconstruct the energy and direction of the original neutrinos. As part of its research and development program, several prototype detector test benches were developed across India. The prototype setup at TIFR, Mumbai is a 12-layer stack of 1 m $\times$ 1 m RPCs without iron layers, designed for characterisation studies and served as a developmental test bench for electronics and data acquisition. The detector also collected cosmic muon data which were used for physics studies and development of data analysis algorithms \cite{samuel2017angular} \cite{samuel2018artificial}. A schematic of the detector setup is shown in Figure \ref{fig:stack}. The data used in this study was collected using this detector setup.

\subsection{Resistive Plate Chambers}
\begin{figure}
    \centering
    \includegraphics[width=0.85\linewidth]{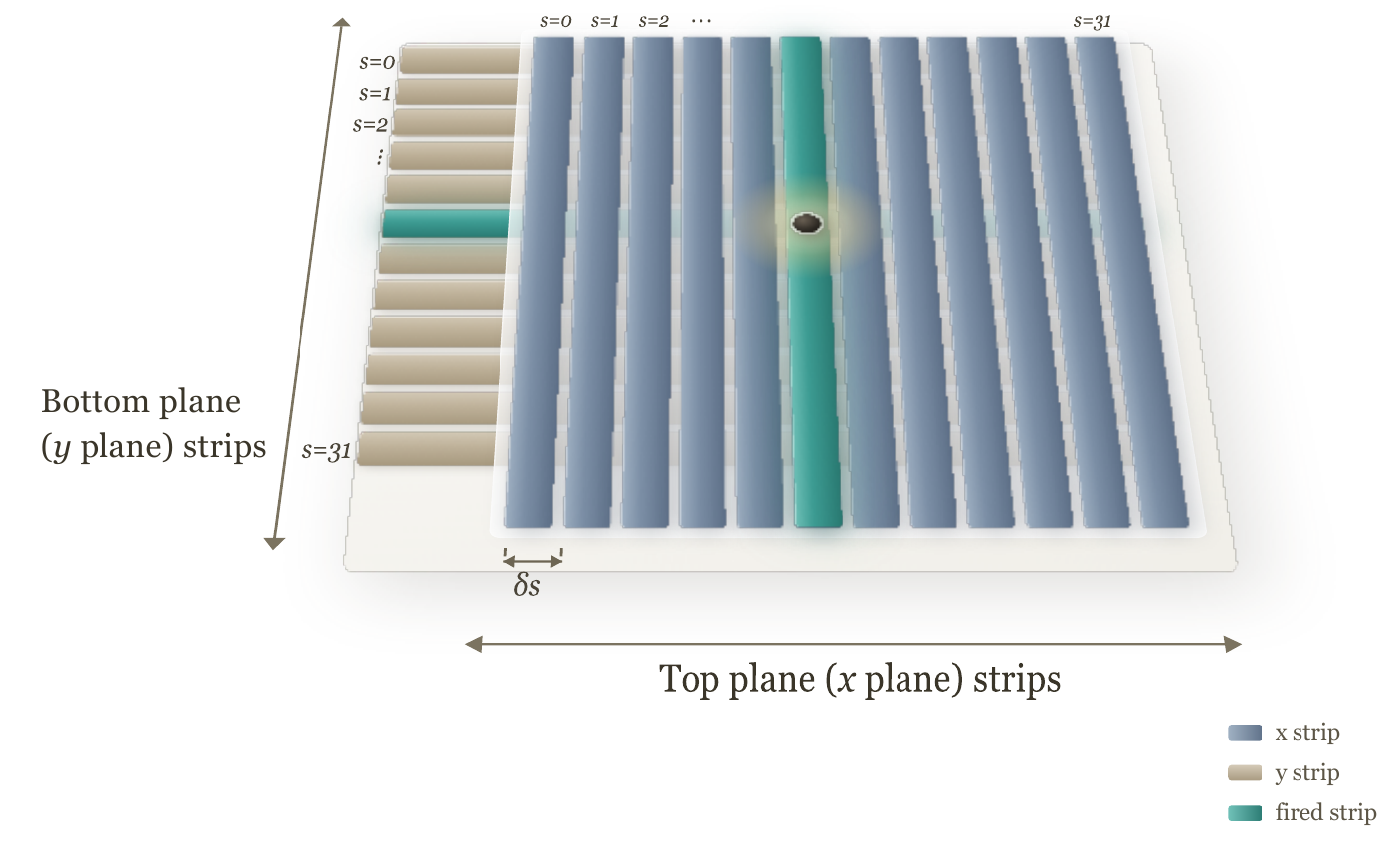}
    \caption{The alignment of strips in the Resistive Plate Chambers at TIFR. Each plane has 32 strips of about 3 cm width and size of the RPC is 1 m $\times$ 1 m.}
    \label{fig:rpc}
\end{figure}

RPCs are gaseous, position-sensitive detectors with a time resolution on the order of a few nanoseconds \cite{datar2009development}. Two glass plates sandwich a gas gap, in which a suitable gas mixture is circulated. The gas mixture acts as the ionising medium.  When a charged particle, such as a muon, enters the gas gap and leads to ionisation, electron-ion pairs are formed. Under the influence of an external electric field applied across the glass plates, the electrons and ions move in opposite directions to their respective electrodes. Copper strips placed on top and bottom of the glass plates detect the electrical signal produced by the motion of electrons and ions in the gas medium. Owing to the localised nature of the ionisation, the specific strip that registers a signal provides a direct measurement of the muon position at a particular moment. The copper strips on the top glass plate are oriented orthogonally to those beneath the bottom glass plate, as shown in Figure \ref{fig:rpc}. Consequently, a muon interaction within a Resistive Plate Chamber (RPC) plane at position $(x,y)$ is identified by the corresponding strip numbers in the two planes, denoted as $(S_x, S_y)$. By stacking multiple RPCs with a defined gap between each, the trajectory of the muon can be tracked, with the $z$ coordinate determined by the position of the RPC along the height of the stack. The RPCs used in the TIFR stack have 32 strips of width 3 cm on either plane with a gap between RPCs ($\delta g$) of approximately 16.8 cm, leading to a stack height, H, of approximately 1.85 m. Additional information regarding the performance characteristics of the detector is available in other studies \cite{datar2009development}.

\subsection{Data acquisition}
The analogue signals from the RPCs are amplified by preamplifiers in the analogue frontend (AFE) with a gain of 80–100 $\times$ and then routed to a  Digital Frontend (DFE) that converts these signals to appropriate digital signals. A VME-based data acquisition system was developed for this stack that reads out the positional and temporal information using Digital IO modules and TDCs (100 ps) of all the RPCs on receipt of a trigger signal. The trigger signal was generated by the coincidence of the signals from the top and bottom RPCs. The trigger rate for this criterion is approximately 10 Hz. The trigger criterion also dictates detector acceptance. In addition to the muon track information, the data acquisition monitored the noise rate of the strips of the RPCs. More details are available in reports dedicated to data acquisition development for the TIFR stack \cite{BHUYAN2012S73}.

\section{Datasets}
The data collected by the RPC stack contain the timing information from each RPCs and the strip numbers from the RPCs that received a $hit$ along both $x$ and $y$ orientations. For a cosmic muon event, the trajectory is a straight line inside the detector volume because multiple coulomb scattering effects are minimal. From previous studies, a detector efficiency of 90\% was observed at optimal operating voltages ($\approx$10 kV). In addition to the \textit{hits} from cosmic muon interactions, \textit{hits} due to detector and electronic noise may also be present. Most studies rely on a simple straight-line fitting to extract physics, such as the zenith angle distribution, after rejecting the noise hits. The dataset used in this study is a collection of files taken between June 2016 and December 2016. A total of 137,632,479 events spread over 45 files were collected over approximately 55 days.


\subsection{Data reduction, event selection}
As the detector setup was primarily used as a test bench, some RPCs may have been noisy or temporarily non-functional. The trigger criteria may have changed during the tests. Therefore, to reduce noise contamination, a preliminary analysis of the datasets was performed, the results of which are summarised here. 
The first analysis is the strip-hit distribution, as shown in Figure \ref{fig:striphits}. 

\begin{figure}
    \centering
    \includegraphics[width=0.75\linewidth]{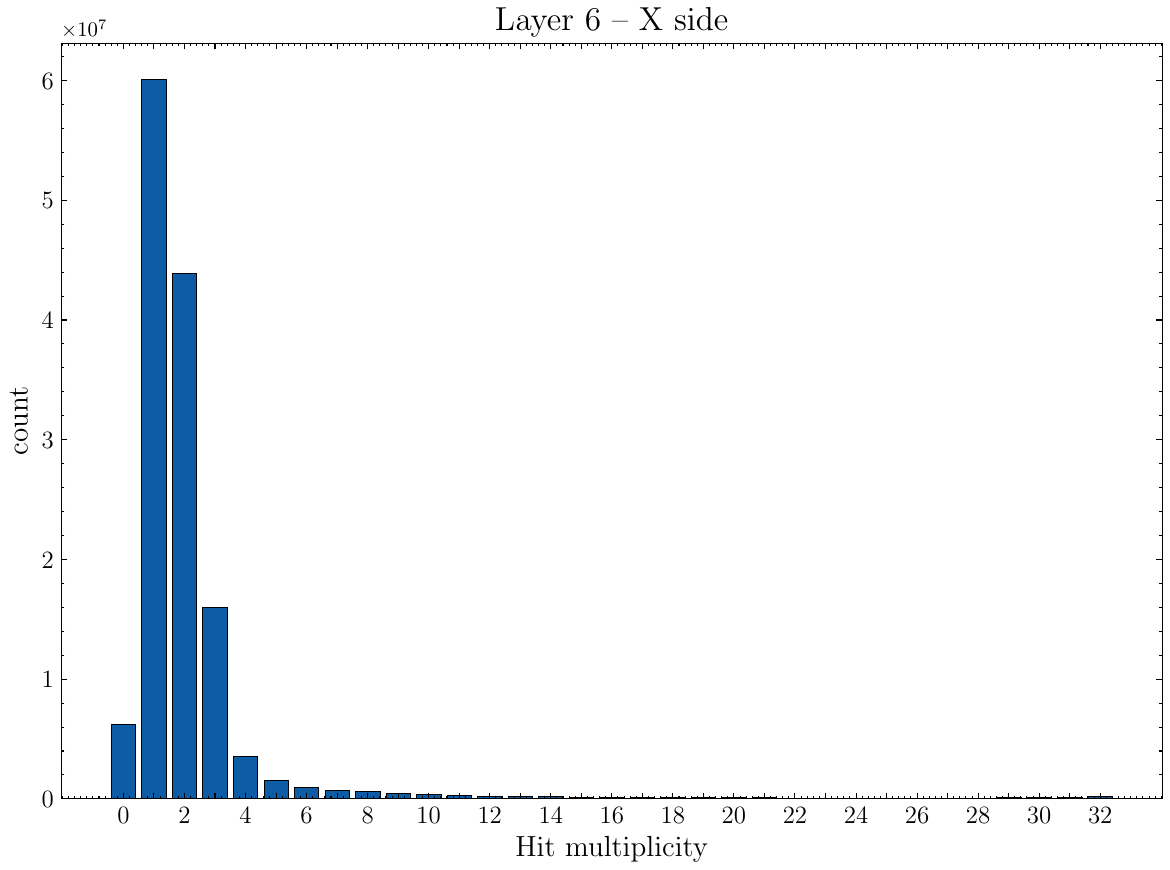}
    \caption{The hit multiplicity for layer 6 shows that while single-hit frequency is more dominant, double-hit frequency is non-negligible.}
    \label{fig:hitmultiplicity}
\end{figure}

The distribution shows that for layers 2-6, the strip hits are fairly concentrated in the central region owing to the acceptance of the detector. Some strips show relatively higher/lower counts (layer 2, strip 31 or layer 5, strip 29). Layer 1 also shows very few counts, as it was non-functional most of the time and hence removed from the downstream analysis. Furthermore, the analysis of the strip hit multiplicity showed that while single-hit patterns were dominant, double-hit patterns also had non-negligible counts. Figure \ref{fig:hitmultiplicity} shows the strip-hit multiplicity distribution for layer 6 on the $x$-side. Based on these observations, the following cuts were applied before proceeding with the fitting procedure:

\begin{figure}
    \centering
    \includegraphics[width=1\linewidth]{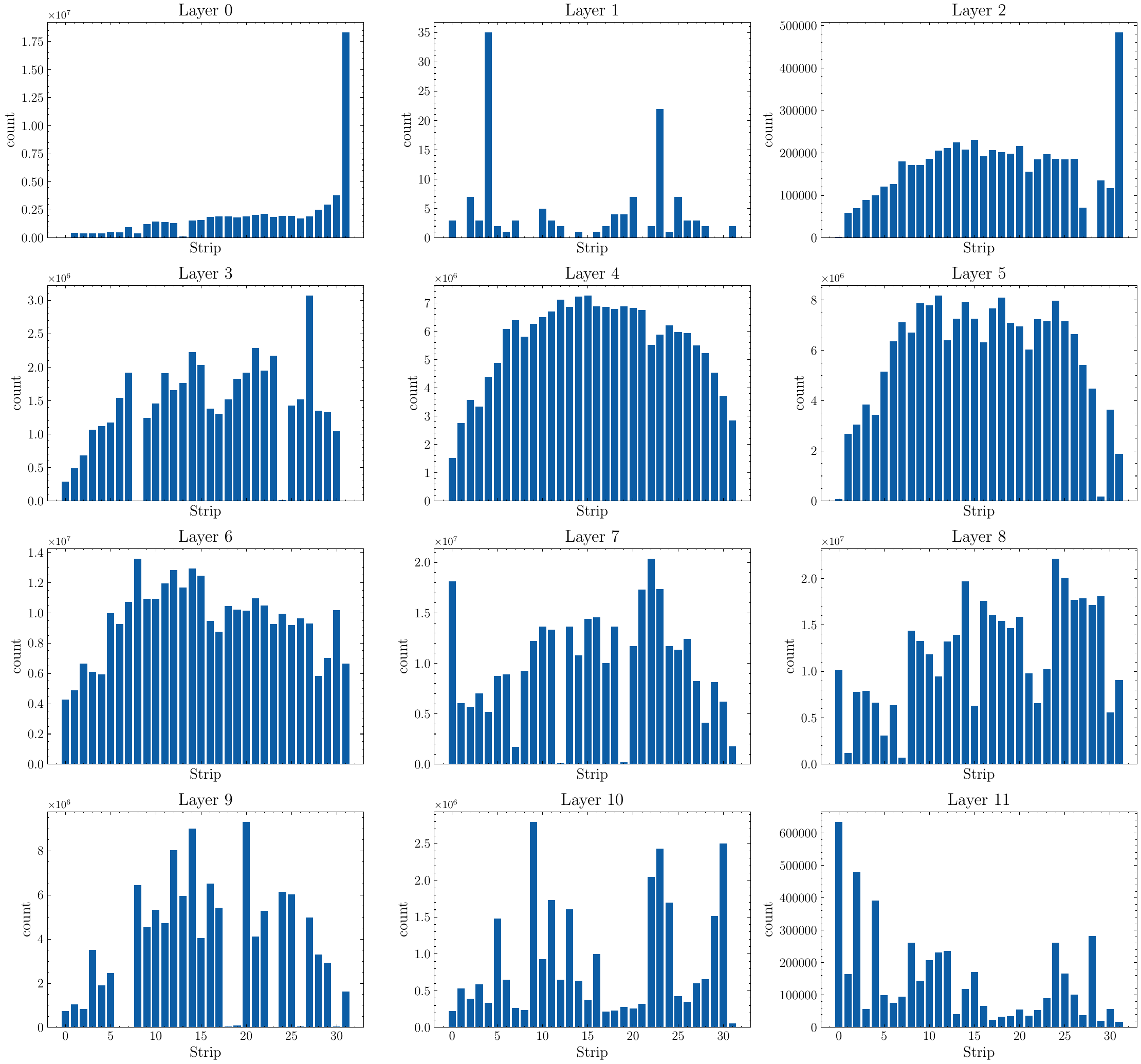}
    \caption{The strip hit distribution(x-side) showing the frequency of counts in each layer. Layer 1 was a non-functional most of the times as indicated by the unusually low strip counts.}
    \label{fig:striphits}
\end{figure}

\begin{enumerate}
    \item Along each projection, only layers with a maximum of 2 hits are considered in the fit.
    \item A minimum of 6 hits are required along each projection.
    \item To reject strips showing unusually high or low counts, strips which are outside $\pm$2-$\sigma$ of the mean counts for the layer are not considered in the fit.
\end{enumerate}
Only approximately 6.6\% of the total events survived this selection cut.

\section{Data analysis}
\begin{figure}
    \centering
    \includegraphics[clip, trim=0.0cm 3.3cm 0.0cm 2.0cm, width=1\linewidth, page=2]{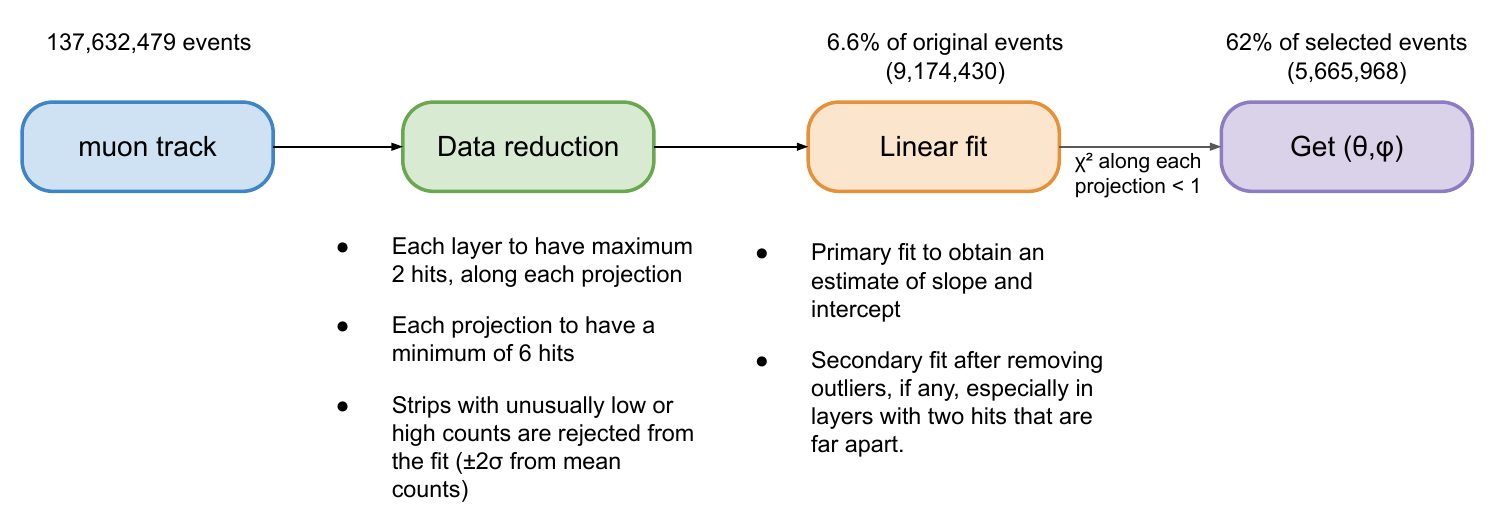}
    \caption{Flowchart showing the data reduction and the data analysis blocks in the process of extracting the zenith and azimuth angles from a muon track.}
    \label{fig:workflow}
\end{figure}

\begin{figure}
    \centering
    \includegraphics[width=1\linewidth]{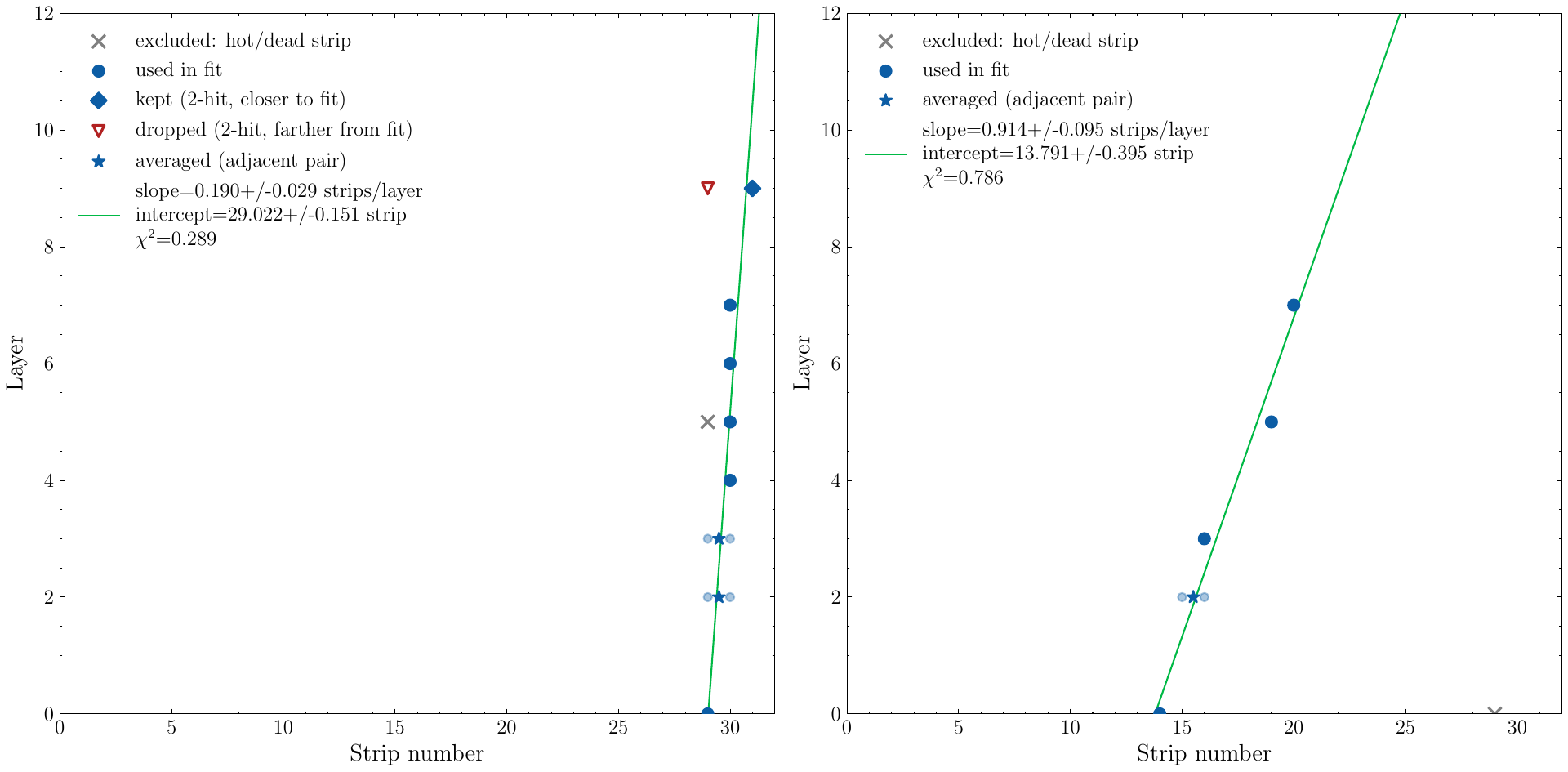}
    \caption{A fit to a sample cosmic muon track along with the hits that pass or fail the selection criteria.}
    \label{fig:sample_event}
\end{figure}

\begin{figure}
    \centering
    \includegraphics[width=1\linewidth]{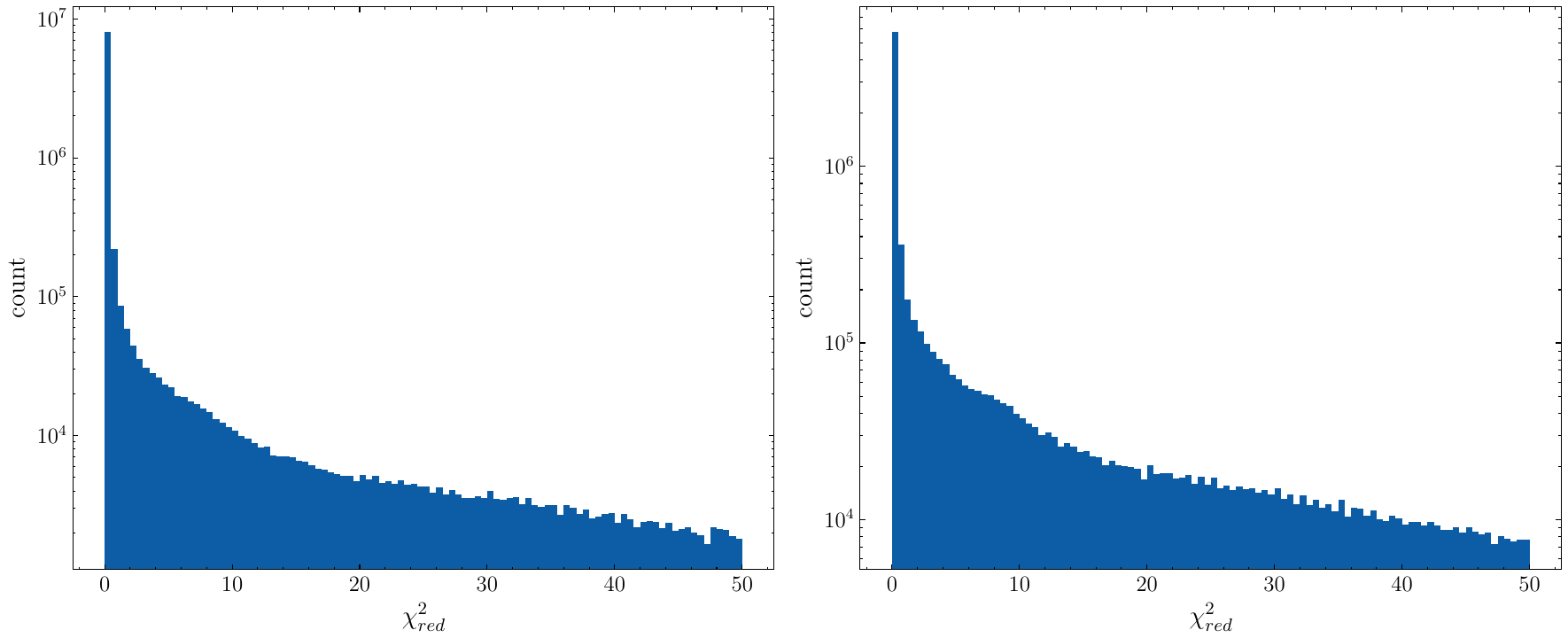}
    \caption{The reduced $\chi^2$ distribution for x-side fit (left) and the y-side fit (right). Only events with $\chi^2<1$ on both the projections were used in the analysis.}
    \label{fig:chisq}
\end{figure}

\begin{figure}
    \centering
    \includegraphics[width=0.65\linewidth]{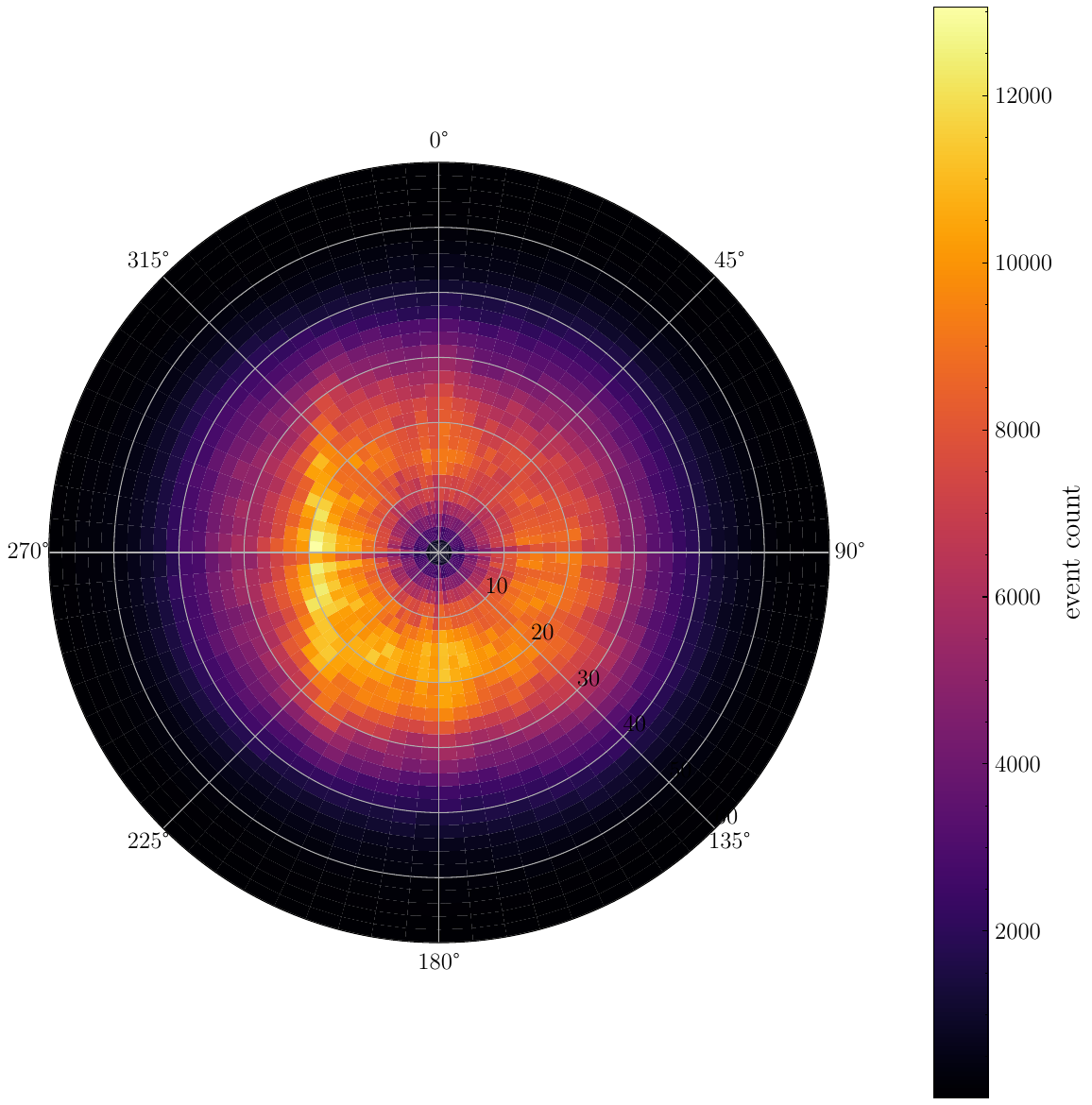}
    \caption{Polar plot of zenith-azimuth distribution. The radius represents the zenith, and the angle represents the azimuth. The slight non-uniformity about the azimuth may be attributed to the structures over the ceiling of the detector room.}
    \label{fig:polar}
\end{figure}

A linear fit was then performed on the events selected according to the criteria mentioned in the previous section. Furthermore, if two hits are adjacent in any layer, the average position is considered the hit position. If not, a primary fit is first performed, the hit far away from the fit line is removed, and a secondary fit is performed on the remaining hits. To further enforce a strict condition on the quality of the fit, only tracks with a reduced $\chi^2$ of $\leq 1$ on both projections were selected for the analysis. Approximately 62\% of the selected events satisfied this condition. A sample cosmic muon event showing the hits that passed or failed  the selection criteria is shown in Figure \ref{fig:sample_event}. The $\chi^2$ distribution from both the projections is shown in Figure \ref{fig:chisq}. The slopes and intercepts from $x$ and $y$ projections,  were used to compute the zenith and azimuthal angles of the tracks according to the following formula:

\begin{align}
    \theta &= tan^{-1} \sqrt{m_x^2 + m_y^2}\\
    \phi &= tan^{-1} \frac{m_x} {m_y}
\end{align}
    
where $m_x$ and $m_y$ are the slopes along the respective projections. A polar plot of the zenith and azimuth distributions is shown in Figure \ref{fig:polar}. The workflow for the data analysis is shown in Figure \ref{fig:workflow}.

\section{Random number generation}
\label{sec:rng}
\begin{figure}
    \centering
    \includegraphics[width=0.5\linewidth]{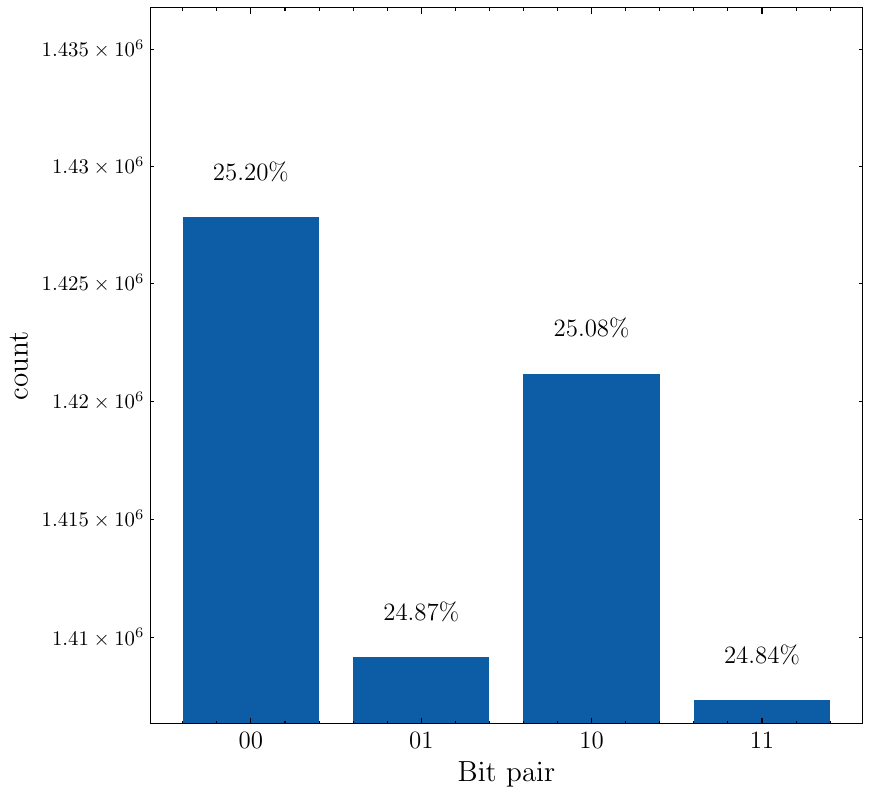}
    \caption{Bit pair distribution based on the parity of the zenith and azimuth angles showing the slight non-uniformity amongst the possible pairs. The azimuth bit is represented by LSB. }
    \label{fig:bitpair}
\end{figure}

As seen in the previous section, for every muon track, two parameters are obtained, namely, the zenith angle $\theta$ and the azimuth angle $\phi$. These two angles encode the muon arrival direction. The arrival directions of cosmic muons are expected to be random in nature, in the sense that the arrival direction of an event does not depend on the previous events. Therefore, several methods can be conceived to generate random numbers from these arrival directions. A simple and direct approach involves binning the two angles and assigning 0 or 1 based on the parity of the bin in which the angles of a track fall. In our study, the zenith angle was binned in 1$^\circ$ bins from 0-60$^\circ$ and the azimuth was binned in 1$^\circ$ bins from -180-180$^\circ$. The two bits are then concatenated with the azimuth bit forming the least significant bit. In this technique, therefore, two bits are generated for each detected muon, the distribution of which is shown in Figure \ref{fig:bitpair}. The plot (in logarithmic scale) reveals a slight non-uniformity among the generated bit pairs. Specifically, the bit pairs in which the azimuth bit has an odd parity seem to be suppressed, indicating a certain asymmetry along $\phi$ in the muon arrival direction. The difference in the odd-even parities for the zenith bit was 0.15\%, whereas for the azimuth bit, it was 0.57\%. This slight disparity may be due to the structure above the ceiling of the detector room or even due to noisy/dead strip combinations in the detectors. This necessitates the use of a conditioning algorithm to remove the bias. We used a simple XOR conditioning  which improved the parity imbalance to 0.08\% at the cost of reducing the bit count to 1 bit/symbol compared to the earlier 2 bits/symbol. The XOR conditioned bit patterns were then validated using NIST SP 800-90B methodology for Independent and Identically Distributed (IID) assessment of our bit patterns. The forementioned technique is referred as the baseline method.

In addition to the above technique, few other variations of random number generation were also tried with the aim of improving the bit rate. 
\subsection{Variations of random number generation methodology}
As an initial test, the zenith and azimuth bin indices were each reduced by taking modulo 4, and the resulting 2-bit patterns were subsequently combined via a bitwise XOR operation to yield a single 2-bit pattern for each muon track. This was tried for few other modulo options, namely 8, 16 and 32 each resulting in 3, 4 and 5 bits per muon track thus enhancing the bitrates proportionately. The datasets generated by this mechanism are referred as m4, m8, m16 and m32. In an exploratory attempt to improve the bitrates, a few more variations were also studied. $\theta$  was binned into 64 bins from 0-60$^\circ$ and $\phi$ into 256 bins from -180-180$^\circ$. The 6 $\theta$ bits representing the zenith angle bin were concatenated with the 8 $\phi$  bits representing the azimuth bin to form a 14-bit pattern for a single muon track. The 14-bit pattern from two tracks are then XOR-ed to create a new 14-bit pattern. This is referred as the cat14 dataset.

\begin{figure}
    \centering
    \includegraphics[clip, trim=0.0cm 0cm 0.0cm 0.0cm, width=1\linewidth, page=1]{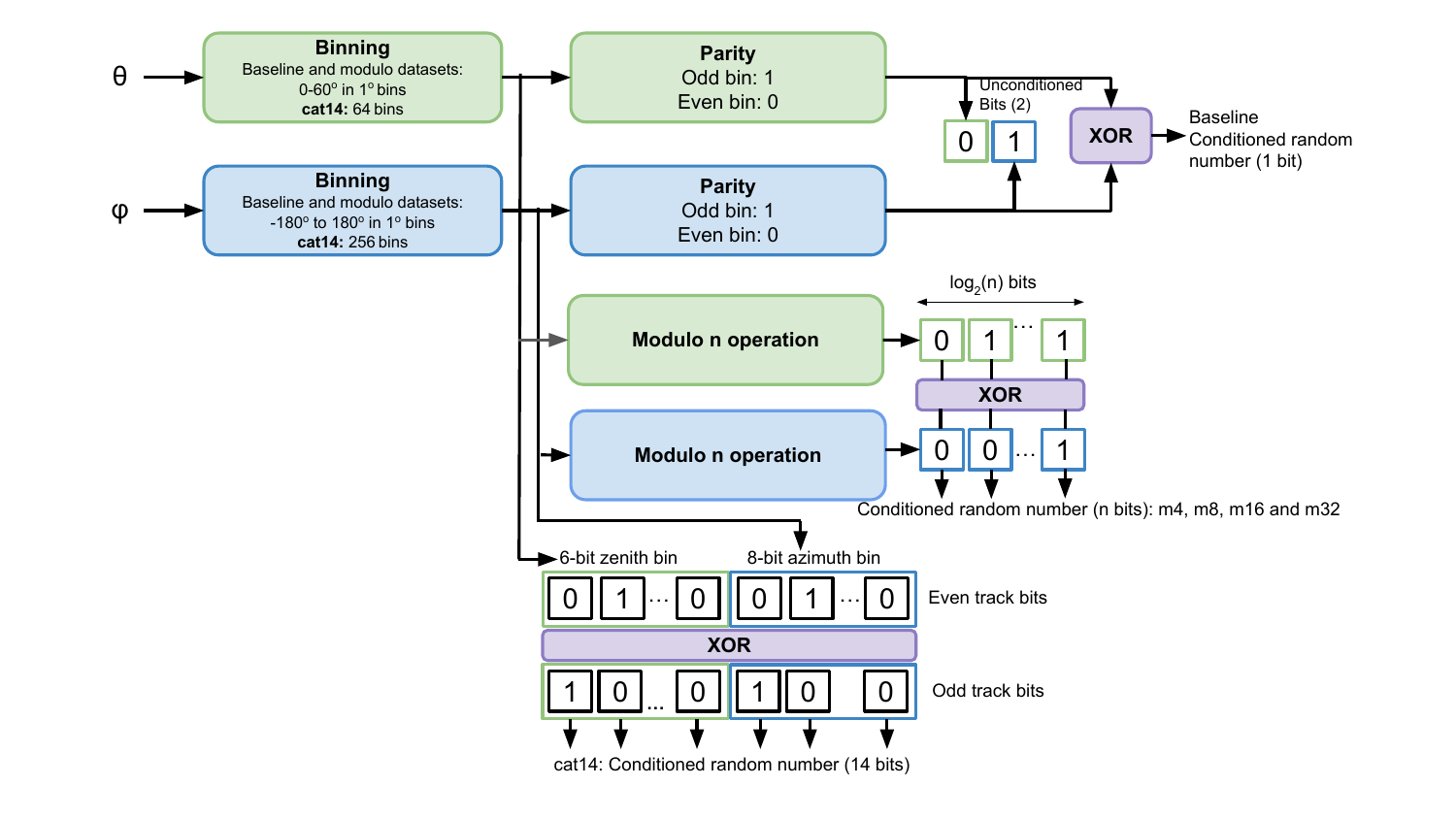}
    \caption{Summary of random number generation techniques proposed in this study.}
    \label{fig:randomgen}
\end{figure}

A block diagram of the random number generation techniques proposed in this study is shown in Figure \ref{fig:randomgen}. The NIST tests were performed on the dataset, and the results are presented in the next section.

\section{Results of NIST SP 800-90B statistical tests}
NIST SP 800-90B specifies a battery of tests for validating entropy sources \cite{sonmez2016recommendation}. The independent and identically distributed (IID) track tests essentially verify if the probability of 0 is similar to the probability of 1 and that the individual bits are independent from each other. For instance, the test for independence splits the data into m-bit tuples and checks how frequently a given m-bit tuple occurs in the dataset and compares with the expected value if the bits are assumed to be independent. Goodness-of-fit test splits the data into non-overlapping samples of length L/10 where L is the number of bits in the full dataset. If the distribution is identical, the number of 0s and 1s in the original dataset should essentially be similar to the distribution in the sliced datasets. In addition to these $\chi^2$ tests, permutation tests shuffle the dataset and compute a test statistic for each of the shuffled patterns and compare it with the test statistic of the original dataset. 19 permutation tests are performed on each of the shuffled patterns. In order to qualify as a TRNG, the dataset generated should pass both the permutation tests and the $\chi^2$ tests. The quality of randomness is quantified by min-entropy which determines the amount of entropy in each sample (i.e, bit). For the IID track, the proportion $\hat{p}$ of the most common bit is used to calculate the upper bound on the probability $p_u$ assuming binomial distribution. The min-entropy is then calculated as:

\begin{equation*}
H_{original} = -log_2(p_u)    
\end{equation*}
A value close to 1 indicates a high quality of randomness. More details of these tests can be found in \cite{sonmez2016recommendation}. The results for the baseline dataset indicate a high quality of randomness with $H_{original}=0.997202$ bits/bit.

The NIST results of the modulo-based datasets m4, m8, m16, and m32 show a $H_{min}$ of 0.998663, 0.998583, 0.999182, and 0.914664 bits/bit, respectively. The m4, m8, and m16 datasets passed the NIST tests, whereas the m32 dataset failed the permutation tests.  Therefore, the maximum bitrate that can be achieved is 4 bits per track using this technique. Assuming a nominal event rate of 1 \text{ cm}$^{-2}$\text{ min}$^{-1}$ for a detector of lateral dimensions 1 m $\times$ 1 m, the average bit rate is approximately 650 bits per second, which can be increased to approximately 2500 bits per second with 2 m $\times$ 2 m RPCs.

\begin{figure}
    \centering
       \includegraphics[clip, trim=0.0cm 0cm 0.0cm 0.0cm, width=1\linewidth, page=4]{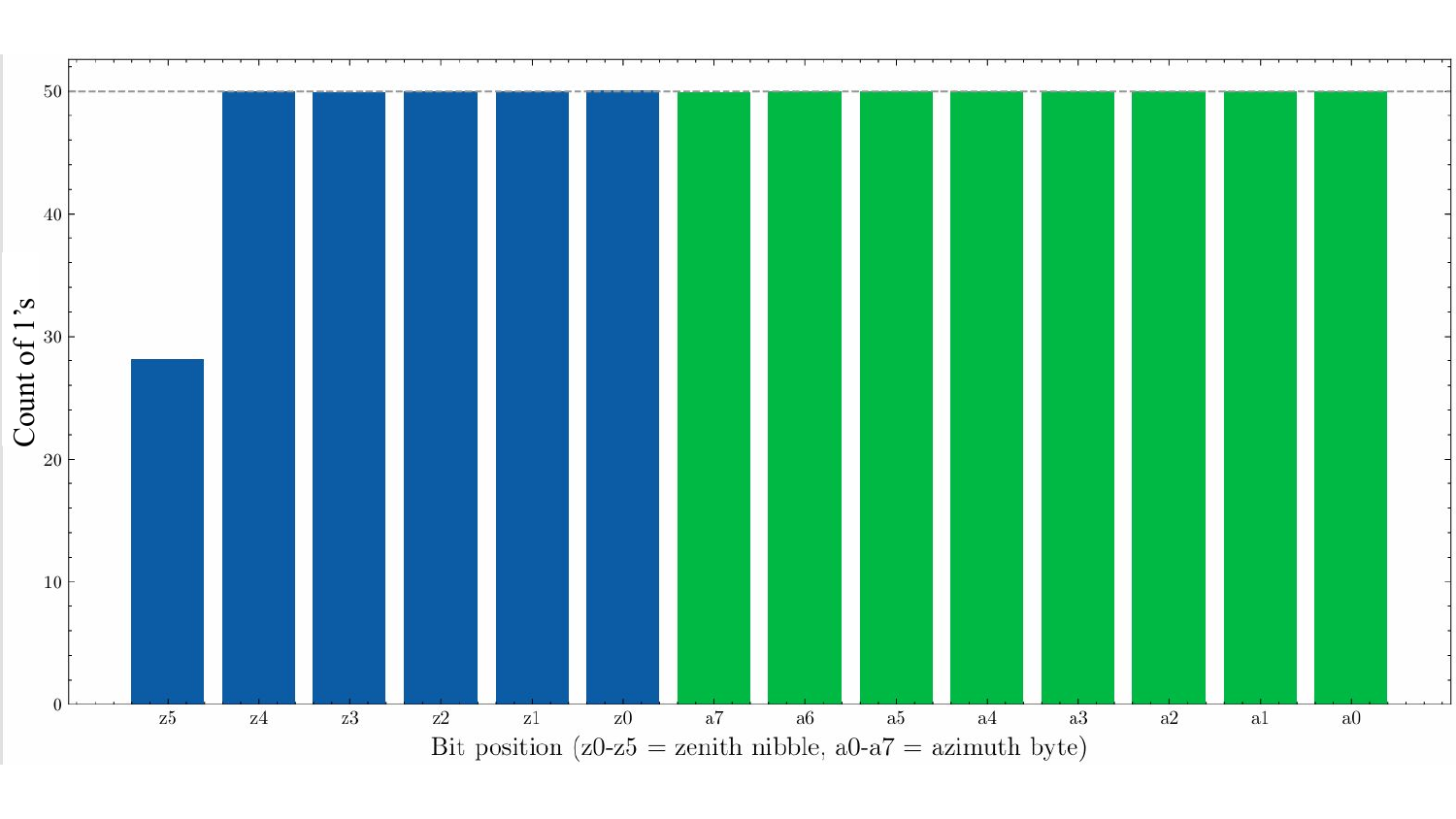}
    \caption{The distribution of 1's in the concatenation style random number generation technique. The MSB distribution is significantly lower than the rest owing to the low counts at large angles. }
    \label{fig:concat-dist}
\end{figure}

The concatenation dataset cat14 showed a non-uniform distribution in the most significant bit (MSB), shown in Figure \ref{fig:concat-dist},  which is attributed to the low counts of muons at large angles, leading to a failure in the $\chi^2$ and permutation tests. The test was repeated by ignoring the MSB, but the dataset still failed the IID tests. The results are summarised in Table \ref{table:comparsion}.

%
%

\begin{table}[htbp]
\centering
\caption{Comparison of NIST SP 800-90B IID-track test results across the variations in the random number generation mechanism. IID perm.\ test is the overall result of the SP 800-90B permutation IID test battery (19 sub-statistics).}
\label{tab:sp90b_comparison}
\resizebox{\textwidth}{!}{%
\begin{tabular}{lrccccccc}
\toprule
Dataset & $N$ (samples) & $H_{\mathrm{original}}$ (bits/bit) & MCV $\hat{p}$ & $\chi^2$ indep.\ $p$-value & $\chi^2$ GoF $p$-value & $\chi^2$ test &  IID perm.\ test \\
\midrule
baseline      & 5,665,567  & 0.997202 & 0.500430 & 0.573560 & 0.200917 & Passed  &  Passed \\
m4                 & 11,331,134 & 0.998663 & 0.500081 & 0.249233 & 0.491546 & Passed  &  Passed \\
m8                 & 16,996,701 & 0.998583 & 0.500179 & 0.132364 & 0.097219 & Passed  &  Passed \\
m16                & 22,662,268 & 0.999182 & 0.500013 & 0.737067 & 0.027252 & Passed  &  Passed \\
m32                & 28,327,835 & 0.998700 & 0.500209 & 0.040182 & 0.006432 & Passed  &  Failed \\
cat14 (no MSB)  & 36,826,179 & 0.998455 & 0.500323 & 0.000011 & 0.812185 & Failed  &  Failed \\
\bottomrule
\end{tabular}%
}
\label{table:comparsion}
\end{table}

\section{Conclusions}
We have presented here, for the first time,  a random number generator based on the arrival direction of cosmic muons and have shown that under specific conditioning, it is possible to achieve upto 4 bits per track with high entropy.  Although the bit rate and detector size presented in this study may not be optimal for practical applications, given that the detector was designed for a different purpose, there remains significant potential for enhancement. The results of this study will be useful as a proof-of-principle. Because the detector was used as a test-bench, data reduction resulted in discarding many events that would have otherwise deteriorated the maximum entropy. Under standard conditions, it is possible to achieve close to 650 bits per second which, in principle, can be improved with RPCs of larger dimension and with better angular resolution. The bit rate also scales with size of the detector and the acceptance. With detectors of better spatial resolution, fine binning of the arrival directions may also lead to improved bit rates. As discussed, the random number generator also has the potential for experiments testing the setting-independance loophole although a detailed investigation is required to understand the feasibility. In the proposal that uses signals from quasars for QRNG, an important factor that had to be considered is the coincidence rates which is the probability that two detectors will simultaneously generate a random number in a given time interval. If the mean cosmic muon rate of 1 \text{ cm}$^{-2}$\text{ min}$^{-1}$ is assumed, with a detector efficiency of 95\%, and detector separation distance taken to be 100 km, the probability that two detector detect at least one muon in the time the entangled particles travel can be calculated using:

\begin{equation}
    P_{simultaneous} = (1-e^{-\epsilon R\Delta t})^2  
\end{equation}
where, $\epsilon$ is the efficiency of the muon detector, $\Delta t$ is the time the entangled photons take to travel from source to detector and R is the rate of cosmic muons. For the values assumed, the probability turns out be 0.2\% which is impractical. However, increasing the RPC size to 5 m $\times$ 5 m, enhances the probability to 48\% which implies a possible use of the setup as a QRNG for closing the setting-independance loophole. 
\section{Acknowledgements}
During the preparation of this work, the author used Paperpal for grammar and language correction, and Claude (Anthropic) to assist in writing analysis code and generating plots/figures. After using these tools, the author reviewed, tested, and edited all outputs as needed, and take full responsibility for the content of the publication, including the validity of the analysis and results.
\printbibliography 

@article{ParticleDataGroup:2026aaa,
    author = "Takahashi, F. and others",
    collaboration = "Particle Data Group",
    title = "{Review of Particle Physics}",
    doi = "10.1142/S0217751X26300115",
    journal = "Int. J. Mod. Phys. A",
    volume = "41",
    pages = "2630011",
    year = "2026"
}

@article{abubakar2026explanation,
  title={Explanation of the seasonal variation of cosmic multiple muon events observed with the NOvA Near Detector},
  author={Abubakar, S and Acero, MA and Acharya, B and Adamson, P and Anfimov, N and Antoshkin, A and Arrieta-Diaz, E and Asquith, Lily and Aurisano, A and Back, A and others},
  journal={Physical Review D},
  volume={113},
  number={1},
  pages={012001},
  year={2026},
  publisher={APS}
}

@article{Kwon:09,
author = {Osung Kwon and Young-Wook Cho and Yoon-Ho Kim},
journal = {Appl. Opt.},
number = {9},
pages = {1774--1778},
publisher = {Optica Publishing Group},
title = {Quantum random number generator using photon-number path entanglement},
volume = {48},
month = {Mar},
year = {2009},
url = {https://opg.optica.org/ao/abstract.cfm?URI=ao-48-9-1774},
doi = {10.1364/AO.48.001774},
}

@article{CACCIA2020164480,
title = {In-silico generation of random bit streams},
journal = {Nuclear Instruments and Methods in Physics Research Section A: Accelerators, Spectrometers, Detectors and Associated Equipment},
volume = {980},
pages = {164480},
year = {2020},
issn = {0168-9002},
doi = {https://doi.org/10.1016/j.nima.2020.164480},
url = {https://www.sciencedirect.com/science/article/pii/S0168900220308779},
author = {M. Caccia and L. Malinverno and L. Paolucci and C. Corridori and E. Proserpio and A. Abba and A. Cusimano and W. Kucewicz and P. Dorosz and M. Baszczyk and M. Esposito and P. Svenda}
}

@INPROCEEDINGS{9159728,
  author={Gamil, Homer and Mehta, Pranav and Chielle, Eduardo and Giovanni, Adriano Di and Nabeel, Mohammed and Arneodo, Francesco and Maniatakos, Michail},
  booktitle={2020 IEEE 26th International Symposium on On-Line Testing and Robust System Design (IOLTS)}, 
  title={Muon-Ra: Quantum random number generation from cosmic rays}, 
  year={2020},
  volume={},
  number={},
  pages={1-6},
  doi={10.1109/IOLTS50870.2020.9159728}}

@techreport{sonmez2016recommendation,
  title={Recommendation for the entropy sources used for random bit generation},
  author={S{\"o}nmez Turan, Meltem and Barker, Elaine and Kelsey, John and McKay, Kerry and Baish, Mary and Boyle, Michael},
  year={2016},
  institution={National Institute of Standards and Technology}
}

@article{ICAL:2015stm,
    author = "Ahmed, Shakeel and others",
    collaboration = "ICAL",
    title = "{Physics Potential of the ICAL detector at the India-based Neutrino Observatory (INO)}",
    eprint = "1505.07380",
    archivePrefix = "arXiv",
    primaryClass = "physics.ins-det",
    reportNumber = "INO-ICAL-PHY-NOTE-2015-01",
    doi = "10.1007/s12043-017-1373-4",
    journal = "Pramana",
    volume = "88",
    number = "5",
    pages = "79",
    year = "2017"
}

@article{BHUYAN2012S73,
title = {VME-based data acquisition system for the India-based Neutrino Observatory prototype detector},
journal = {Nuclear Instruments and Methods in Physics Research Section A: Accelerators, Spectrometers, Detectors and Associated Equipment},
volume = {661},
pages = {S73-S76},
year = {2012},
note = {X. Workshop on Resistive Plate Chambers and Related Detectors (RPC 2010)},
issn = {0168-9002},
doi = {https://doi.org/10.1016/j.nima.2010.08.075},
url = {https://www.sciencedirect.com/science/article/pii/S0168900210018498},
author = {M. Bhuyan and V.B. Chandratre and S. Dasgupta and V.M. Datar and S.D. Kalmani and S.M. Lahamge and N.K. Mondal and P. Nagaraj and S. Pal and S.K. Rao and A. Redij and D. Samuel and M.N. Saraf and B. Satyanarayana and R.R. Shinde and S.S. Upadhya}
}

@article{PhysRevLett.81.1562,
  title = {Evidence for Oscillation of Atmospheric Neutrinos},
  author = {Fukuda, Y. and Hayakawa, T. and Ichihara, E. and Inoue, K. and Ishihara, K. and Ishino, H. and Itow, Y. and Kajita, T. and Kameda, J. and Kasuga, S. and Kobayashi, K. and Kobayashi, Y. and Koshio, Y. and Miura, M. and Nakahata, M. and Nakayama, S. and Okada, A. and Okumura, K. and Sakurai, N. and Shiozawa, M. and Suzuki, Y. and Takeuchi, Y. and Totsuka, Y. and Yamada, S. and Earl, M. and Habig, A. and Kearns, E. and Messier, M. D. and Scholberg, K. and Stone, J. L. and Sulak, L. R. and Walter, C. W. and Goldhaber, M. and Barszczxak, T. and Casper, D. and Gajewski, W. and Halverson, P. G. and Hsu, J. and Kropp, W. R. and Price, L. R. and Reines, F. and Smy, M. and Sobel, H. W. and Vagins, M. R. and Ganezer, K. S. and Keig, W. E. and Ellsworth, R. W. and Tasaka, S. and Flanagan, J. W. and Kibayashi, A. and Learned, J. G. and Matsuno, S. and Stenger, V. J. and Takemori, D. and Ishii, T. and Kanzaki, J. and Kobayashi, T. and Mine, S. and Nakamura, K. and Nishikawa, K. and Oyama, Y. and Sakai, A. and Sakuda, M. and Sasaki, O. and Echigo, S. and Kohama, M. and Suzuki, A. T. and Haines, T. J. and Blaufuss, E. and Kim, B. K. and Sanford, R. and Svoboda, R. and Chen, M. L. and Conner, Z. and Goodman, J. A. and Sullivan, G. W. and Hill, J. and Jung, C. K. and Martens, K. and Mauger, C. and McGrew, C. and Sharkey, E. and Viren, B. and Yanagisawa, C. and Doki, W. and Miyano, K. and Okazawa, H. and Saji, C. and Takahata, M. and Nagashima, Y. and Takita, M. and Yamaguchi, T. and Yoshida, M. and Kim, S. B. and Etoh, M. and Fujita, K. and Hasegawa, A. and Hasegawa, T. and Hatakeyama, S. and Iwamoto, T. and Koga, M. and Maruyama, T. and Ogawa, H. and Shirai, J. and Suzuki, A. and Tsushima, F. and Koshiba, M. and Nemoto, M. and Nishijima, K. and Futagami, T. and Hayato, Y. and Kanaya, Y. and Kaneyuki, K. and Watanabe, Y. and Kielczewska, D. and Doyle, R. A. and George, J. S. and Stachyra, A. L. and Wai, L. L. and Wilkes, R. J. and Young, K. K.},
  collaboration = {Super-Kamiokande Collaboration},
  journal = {Phys. Rev. Lett.},
  volume = {81},
  issue = {8},
  pages = {1562--1567},
  numpages = {0},
  year = {1998},
  month = {Aug},
  publisher = {American Physical Society},
  doi = {10.1103/PhysRevLett.81.1562},
  url = {https://link.aps.org/doi/10.1103/PhysRevLett.81.1562}
}

@article{GUPTA2005311,
title = {GRAPES-3—A high-density air shower array for studies on the structure in the cosmic-ray energy spectrum near the knee},
journal = {Nuclear Instruments and Methods in Physics Research Section A: Accelerators, Spectrometers, Detectors and Associated Equipment},
volume = {540},
number = {2},
pages = {311-323},
year = {2005},
issn = {0168-9002},
doi = {https://doi.org/10.1016/j.nima.2004.11.025},
url = {https://www.sciencedirect.com/science/article/pii/S016890020402426X},
author = {S.K. Gupta and Y. Aikawa and N.V. Gopalakrishnan and Y. Hayashi and N. Ikeda and N. Ito and A. Jain and A.V. John and S. Karthikeyan and S. Kawakami and T. Matsuyama and D.K. Mohanty and P.K. Mohanty and S.D. Morris and T. Nonaka and A. Oshima and B.S. Rao and K.C. Ravindran and M. Sasano and K. Sivaprasad and B.V. Sreekantan and H. Tanaka and S.C. Tonwar and K. Viswanathan and T. Yoshikoshi}
}

@article{kaiser2020tackling,
  title={Tackling Loopholes in Experimental Tests of Bell's Inequality},
  author={Kaiser, David I},
  journal={arXiv preprint arXiv:2011.09296},
  year={2020}
}

@article{PhysRevLett.112.110405,
  title = {Testing Bell's Inequality with Cosmic Photons: Closing the Setting-Independence Loophole},
  author = {Gallicchio, Jason and Friedman, Andrew S. and Kaiser, David I.},
  journal = {Phys. Rev. Lett.},
  volume = {112},
  issue = {11},
  pages = {110405},
  numpages = {5},
  year = {2014},
  month = {Mar},
  publisher = {American Physical Society},
  doi = {10.1103/PhysRevLett.112.110405},
  url = {https://link.aps.org/doi/10.1103/PhysRevLett.112.110405}
}

@article{samuel2017angular,
  title={Angular resolution of stacked resistive plate chambers},
  author={Samuel, Deepak and Onikeri, Pratibha B and Murgod, Lakshmi P},
  journal={Journal of Cosmology and Astroparticle Physics},
  volume={2017},
  number={01},
  pages={058--058},
  year={2017}
}

@article{datar2009development,
  title={Development of glass resistive plate chambers for INO experiment},
  author={Datar, VM and Jena, Satyajit and Kalmani, SD and Mondal, NK and Nagaraj, P and Reddy, LV and Saraf, M and Satyanarayana, B and Shinde, RR and Verma, P},
  journal={Nuclear Instruments and Methods in Physics Research Section A: Accelerators, Spectrometers, Detectors and Associated Equipment},
  volume={602},
  number={3},
  pages={744--748},
  year={2009},
  publisher={Elsevier}
}

@article{samuel2018artificial,
  title={Artificial neural networks-based track fitting of cosmic muons through stacked resistive plate chambers},
  author={Samuel, Deepak and Suresh, Karthik},
  journal={Journal of Instrumentation},
  volume={13},
  number={10},
  pages={P10035--P10035},
  year={2018}
}

@article{PhysRevE.72.016220,
  title = {Properties making a chaotic system a good pseudo random number generator},
  author = {Falcioni, Massimo and Palatella, Luigi and Pigolotti, Simone and Vulpiani, Angelo},
  journal = {Phys. Rev. E},
  volume = {72},
  issue = {1},
  pages = {016220},
  numpages = {10},
  year = {2005},
  month = {Jul},
  publisher = {American Physical Society},
  doi = {10.1103/PhysRevE.72.016220},
  url = {https://link.aps.org/doi/10.1103/PhysRevE.72.016220}
}

@article{shor1999polynomial,
  title={Polynomial-time algorithms for prime factorization and discrete logarithms on a quantum computer},
  author={Shor, Peter W},
  journal={SIAM review},
  volume={41},
  number={2},
  pages={303--332},
  year={1999},
  publisher={SIAM}
}

@book{chen2016report,
  title={Report on post-quantum cryptography},
  author={Chen, Lily and Chen, Lily and Jordan, Stephen and Liu, Yi-Kai and Moody, Dustin and Peralta, Rene and Perlner, Ray A and Smith-Tone, Daniel},
  volume={12},
  year={2016},
  publisher={US Department of Commerce, National Institute of Standards and Technology~…}
}
\end{document}